\documentclass[twocolumn]{aastex63}
\usepackage{amsmath}
\usepackage{color}
\usepackage{ulem}

\begin{document}
\newcommand{\li}[1]{{\color{blue} #1}}
\newcommand{\hao}[1]{{\color{red} #1}}
\newcommand{\lyr}{Ly$\gamma$}
\newcommand{\lyb}{Ly$\beta$}
\newcommand{\lya}{Ly$\alpha$}
\newcommand{\hr}{H$\gamma$}
\newcommand{\hb}{H$\beta$}
\newcommand{\ha}{H$\alpha$}
\newcommand{\pa}{Pa$\alpha$}
\newcommand{\pb}{Pa$\beta$}
\newcommand{\pr}{Pa$\gamma$}
\newcommand{\ciii}{C\,{\scriptsize III}}
\newcommand{\civ}{C\,{\scriptsize IV}}
\newcommand{\niii}{N\,{\scriptsize III}}
\newcommand{\nv}{N\,{\scriptsize V}}
\newcommand{\oiii}{O\,{\scriptsize III}}
\newcommand{\oiv}{O\,{\scriptsize IV}}
\newcommand{\ovi}{O\,{\scriptsize VI}}
\newcommand{\sii}{S\,{\scriptsize II}}
\newcommand{\siv}{S\,{\scriptsize IV}}
\newcommand{\siiv}{Si\,{\scriptsize IV}}
\newcommand{\feii}{Fe\,{\scriptsize II}}
\newcommand{\kmps}{$\rm km~s^{-1}$}
\newcommand{\mbh}{$M_{\rm BH}$}
\newcommand{\msun}{$M_{\odot}$}

\title{Ultradense Gases beyond Dusty Torus in a Partially Obscured Quasar}

\correspondingauthor{Zhenzhen Li}
\email{lizhenzhen@shao.ac.cn; haol@shao.ac.cn}

\author{Zhenzhen Li}
\affiliation{Key Laboratory for Research in Galaxies and Cosmology, Shanghai Astronomical Observatory, Chinese Academy of Sciences, 80 Nandan Road, Shanghai 200030, China}

\author{Hongyan Zhou}
\affiliation{Polar Research Institute of China, Jinqiao Rd. 451, Shanghai, 200136, China}
\affiliation{Department of Astronomy, University of Science and Technology of China, Hefei, Anhui 230026, China}

\author{Lei Hao}
\affiliation{Key Laboratory for Research in Galaxies and Cosmology, Shanghai Astronomical Observatory, Chinese Academy of Sciences, 80 Nandan Road, Shanghai 200030, China}

\author{Xiheng Shi}
\affiliation{Polar Research Institute of China, Jinqiao Rd. 451, Shanghai, 200136, China}
\affiliation{Department of Astronomy, University of Science and Technology of China, Hefei, Anhui 230026, China}

\begin{abstract}
  The co-evolution between black holes and galaxies suggests that feedback of active galactic nuclei influence host galaxies through ejecting radiative and kinetic energies to surroundings. Larger scale outflow in local universe are frequently observed by spatially resolved spectroscopy, while smaller scale outflow cannot be directly resolved by current observations. At the scale of the dusty torus, radiative and kinetic energies ejected from the central active nucleus interact with the materials. However, observations of such outflow are rarely reported due to the lack detection of unambiguously gas emission. Here we report the detection of clear and rich emission lines origin from the scale of dusty tours in an partially obscured quasar. The lines share a common intermediate width with full width at half maximum about 1900 \kmps\ and are shown in two systems: a major system is unshifted and a minor system has a blue-shifts of about 2600 \kmps. The line intensity ratios, combining photo-ionization simulations, indicates an ultradense line-emitting region with the density as high as $\sim$ $10^{13}~\rm cm^{-3}$. We interpret this as the lines being excited by a shock induced by the high-density and high-temperature gases at the scale of dusty torus, rather than photo-ionized by the central accretion disk. We speculate that the outflow, launched from the accretion disk, collides onto the inner wall of the dusty torus and shock-heat the gases to cause the major emission lines. The outflowing gases may also collide onto surrounding isolated clouds, and give rise to blue-shifted minor emission lines.
\end{abstract}

\section{INTRODUCTION}
It is well established that massive galaxies generally contain supermassive black holes (SMBHs) in their centers. Observations over the past decades have revealed tight correlations between the SMBH mass and various properties of their host galaxy, such as the velocity dispersion, luminosity and mass of the bulge (e.g., Magorrian et al. 1998; Gebhardt et al. 2000; Merritt \& Ferrarese 2001; McLure \& Dunlop 2002; Tremaine et al. 2002; H{\"a}ring \& Rix 2004; Ferrarese \& Ford 2005; Aller \& Richstone 2007; Gultekin et al. 2009; Woo et al. 2010). It is generally believed that the required link in the co-evolution of SMBH$-$galaxy is provided by the active galactic nuclei (AGNs) feedback (e.g., Silk \& Rees 1998; Fabian 1999; King 2003, 2010; Murray et al. 2005; Fabian 2012). The central black holes influence their entire host galaxies through ejecting kinetic and radiative energies (such as wind, radio jet and radiation pressure) to their surroundings: from the compacted center to extend interstellar and intergalactic media (ISM/IGM). Large scale outflow in local universe are frequently observed by spatially resolved spectroscopy for both ionized (Holt et al., 2008; Fu \& Stockton 2009; Rupke \& Veilleux 2011; Westmoquette et al. 2011) and molecular (Feruglio et al. 2010; Alatalo et al. 2011) gases. The small scale outflow is hard to be directly resolved by current observational techniques and instruments. However, some special emission line features can be quite revealing to provide us opportunities to peek into the compact central region that can be very close to the SMBH (e.g., Zhou et al. 2009).

Here in this paper, we show another example of an QSO presenting unusual emission-line properties and may reveal activities that happen close to its dusty torus region. The QSO show intermediate emission-line widths. The widths of emission lines in QSOs are important properties in the context of the unified frame of AGNs (e.g., Antonucci 1993), which successfully explains the diversity of observed AGNs by assuming a dusty torus located somewhere between the broad emission line region (BELR) and the narrow emission line region (NELR). While it is natural to conceive that there are ``intermediate-emission-line regions (IELR)'' between the BELR and the NELR, the identification of the IELR has been elusive, despite a few previous researches (e.g., Wills et al. 1992; Brotherton et al. 1994; Mason et al. 1996; Crenshaw \& Kraemer 2007; Crenshaw et al. 2009; Hu et al. 2008; Zhu 2009). This is because the emission lines originated from these intermediate regions are not as broad as the broad emission lines, and their signals are easily submerged by the broad emission lines and their detection may rely on delicate and often heavily-degenerated line modeling (Sulentic \& Marziani 1999; Gon{\c{c}}alves et al. 1999). However, identifying these emission lines can be important. Being emitted from different regions as the BLR and NLR, these intermediate-width emission lines, if can be properly identified, can help us to reveal the physics in these regions that can not be covered by studying the broad or narrow emission lines. Recently, Li et al. (2015) reported clear detection of intermediate-width emission lines (IELs) in a partially obscured quasar OI 287 in the \textit{Hubble Space Telescope} (\textit{HST}) Faint Object Spectrograph (FOS) ultraviolet (UV) spectrum. With the BELR obscured by dust and the BELs heavily suppressed, the intermediate emission lines are directly ``seen''. This detection provide a novel method for detecting unambiguous IELs: looking for QSOs that have their broad emission lines, and often the continuum as well, that have been suppressed.

Good UV spectral data can be very useful in using this method to identify this kind of partially-obscured quasars, as the suppression is more easily found in the blue and UV bands where the dust extinction is more significant. The \textit{HST} Space Telescope Imaging Spectrograph (STIS), with a wide wavelength range starting from the far-UV (FUV) band of 1150\AA, provides an opportunity to obtain more clear UV IELs in partially obscured quasars. In this paper, we report a partially obscured quasar 2MASS J15165323+1900482 (referred to as J1516+1900 hereafter). Unlike OI 287, which only shows two strong IELs of \lya\ and \civ\ due to limited observational wavelength range of its \textit{HST} FOS spectrum, J1516+1900 shows rich and clear IELs in the UV spectrum of \textit{HST} STIS. With the aid of these observed information of IELs and by combining the theoretical tool of the photo-ionization simulation, we analyse the physical condition of the IELR in detail, and discuss the origin of the IELs.

This paper is organized as follows. In Section 2, we describe the observations and data reduction; in Section 3, we analyze the observational properties of broadband spectral energy distributions (SED), emission line spectra and dust extinction; in Section 4, we explore the physical conditions of the IELR, and discuss the origin of the IELs; finally, we give a brief summary in Section 5. Throughout this paper, we use the cosmological parameters $H_0 = 70 ~ \rm km~s^{-1} \rm Mpc^{-1}$, $\Omega_{\rm M} = 0.3$, and $\Omega_{\Lambda} = 0.7$.
\section{OBSERVATIONS AND DATA REDUCTION}
\subsection{Photometric Data}
We collected broadband photometric data of J1516+1900 from archives of various large sky surveys, which includes the Wide-field Infrared Survey Explorer (WISE, Wright et al. 2010), the Two Micron All Sky Survey (2MASS; Skrutskie et al. 2006), the Sloan Digital Sky Survey (SDSS, York et al. 2000), the \textit{Galaxy Evolution Explorer} (\textit{GALEX}, Morrissey et al. 2007), and the \textit{Chandra} X-ray Observatory (Wilkes et al. 2002). The details of the multi-wavelength photometric data are presented in Table 1.

\begin{deluxetable}{lrcc}[!htb]
\tabletypesize{\normalsize}
\tablewidth{0pt}
\tablecaption{Photometric Data}
\tablehead{\colhead {Band} & {Value} & {Facility} & {Observed Date} \\
                    {}     & {(mag)} & {}         & {(UT)} }
\startdata
FUV  & 18.494$\pm$0.078 & GALEX  & 2007-04-13 \\
NUV  & 18.049$\pm$0.036 & GALEX  & 2007-04-13 \\
$u$  & 16.198$\pm$0.015 & SDSS   & 2005-04-09 \\
$g$  & 15.614$\pm$0.015 & SDSS   & 2005-04-09 \\
$r$  & 15.169$\pm$0.014 & SDSS   & 2005-04-09 \\
$i$  & 14.592$\pm$0.014 & SDSS   & 2005-04-09 \\
$z$  & 14.861$\pm$0.019 & SDSS   & 2005-04-09 \\
$J$  & 13.437$\pm$0.022 & 2MASS  & 1997-06-15 \\
$H$  & 12.613$\pm$0.022 & 2MASS  & 1997-06-15 \\
$K_s$& 11.376$\pm$0.018 & 2MASS  & 1997-06-15 \\
$W1$ & 10.017$\pm$0.022 & WISE   & 2010-01-30 \\
$W2$ &  8.869$\pm$0.020 & WISE   & 2010-01-30 \\
$W3$ &  6.353$\pm$0.015 & WISE   & 2010-01-30 \\
$W4$ &  4.424$\pm$0.026 & WISE   & 2010-01-30 \\
\enddata
\end{deluxetable}

\subsection{Spectral Data}

\begin{deluxetable*}{cclclc}[!htb]
\tabletypesize{\small}
\tablewidth{0pt}
\tablecaption{Spectroscopic Data}
\tablehead{\colhead {Wavelength Range} & {Slit Width} & {$\lambda/\Delta\lambda$} & {Exp.Time}  & {Instrument}            & {Data}         \\
           \colhead {(\AA)}            & {($''$)}     & {}                        & {(s)}       & {}                      & {(UT)}         }
\startdata
           1150$-$3180                 & 0.5          &  1200, 755                &2857, 2311  & \textit{HST}/STIS       & 2002$-$02$-$11 \\
           3433$-$7841                 & 1.5          &  2195, 1385               &600$\times$2& Shane/DoubleSpec        & 2016$-$07$-$04 \\
           2500$-$10700                & 1.5          &  800, 1700                &600$\times$2& Hale/DoubleSpec         & 2017$-$01$-$09 \\
           9720$-$24629                & 1.5          &  3500                     &300$\times$4& Hale/TripleSpec         & 2016$-$04$-$20 \\
\enddata
\end{deluxetable*}

The spectral data of J1516+1900 cover from FUV to near-infrared (NIR). Table 2 summarizes these observations and we describe the details below.

\emph{HST UV Spectroscopy.} The UV spectrophotometry of J1516+1900 was obtained on 2002 February 11 with the STIS on board the \textit{HST} (PI: Paul Smith, Program ID: 9161). Two spectral observations was taken using a long slit of $52'' \times 0.5''$. One was performed using the G140L grating for a 2857 s exposures, giving a FUV spectrum with a wavelength coverage from 1150 \AA\ to 1730 \AA; the other was obtained using the G230L grating for a 2857 s exposures, producing a near-UV (NUV) spectrum from 1570 \AA\ to 3180 \AA. The data were reduced and calibrated with the \textit{HST} STIS pipeline. We retrieved the spectra from Mikulski Archive for Space Telescopes (MAST)\footnote{http://archive.stsci.edu/}.

\emph{Hale NIR Spectroscopy.} On April 20, 2016, we performed follow-up NIR spectroscopic observations of J1516+1900 using TripleSpec (Wilson et al. 2004) on the Hale Telescope . A slit of $1.1''$ was chosen and four 300 s exposures were taken in an A-B-B-A dithering mode with the primary configuration of the instrument. This yield a spectrum with a wavelength range of $\lambda \sim$ 0.97--2.46 $\mu$m. The data were reduced with the \textit{Triplespectool} package, a modified version of \textit{Spextool} (Cushing et al. 2004).

\emph{Shane/Hale Optical Spectroscopy.} To acquire the optical spectrum, we performed two spectroscopic observations. On July 04, 2016, we took observations of this object using the Kast Double Spectrograph on the Shane Telescope at LICK observatory. A $1.5 ''$ slit was chosen to match the seeing, and two 600 s exposures were taken using the 600 lines mm$^{-1}$ gratings, one blazed at 4310 \AA\ and the other at 5000 \AA. These settings yield a blue wavelength coverages of $\lambda \sim$ 3433--5510 \AA\ and a red coverages of $\lambda \sim$ 5088--7841 \AA, respectively. On January 09 2017, the DoubleSpec on Hale Telescope at Palomar Observatory was used for observation. A $1.5''$ slit was chosen and two 600 s exposures were taken, one used the 600 lines mm$^{-1}$ grating blazed at 4000 \AA\ and the other used the 316 lines mm$^{-1}$ grating blazed at 7500 \AA. This gave a spectrum with a wavelength range from 3000 \AA\ to 5700 \AA\ in blue, and a spectrum from 4800 \AA\ to 10700 \AA\ in red. Standard stars were observed quasi-simultaneously for flux calibration. Wavelength calibration was carried out using an Fe/Ar lamp for the blue portion and He/Ne/Ar lamp for the red portion. The data reduction was accomplished with standard procedures using IRAF\footnote{IRAF is distributed by the National Optical Astronomy Observatory, which is operated by the Association of Universities for Research in Astronomy, Inc., under cooperative agreement with the National Science Foundation.}. We combined the DoubleSpec and SDSS spectrum to form one spectrum covering a wavelength range of $\lambda \sim$ 3000$-$10700 \AA.

Before conducting analysis, all of the photometric and spectroscopic data have been corrected for a Galactic reddening of $E(B-V)$=0.040 using the updated dust map of Schlafly \& Finkbeiner (2011), and converted to the rest frame of the quasar using the redshift $z=0.1893$ determined by the peak of [\oiii]~$\lambda$5007 emission line.

\section{DATA ANALYSIS and RESULTS}
\subsection{Broadband SED}

\begin{figure*}[!htb]
\epsscale{0.8}
\plotone{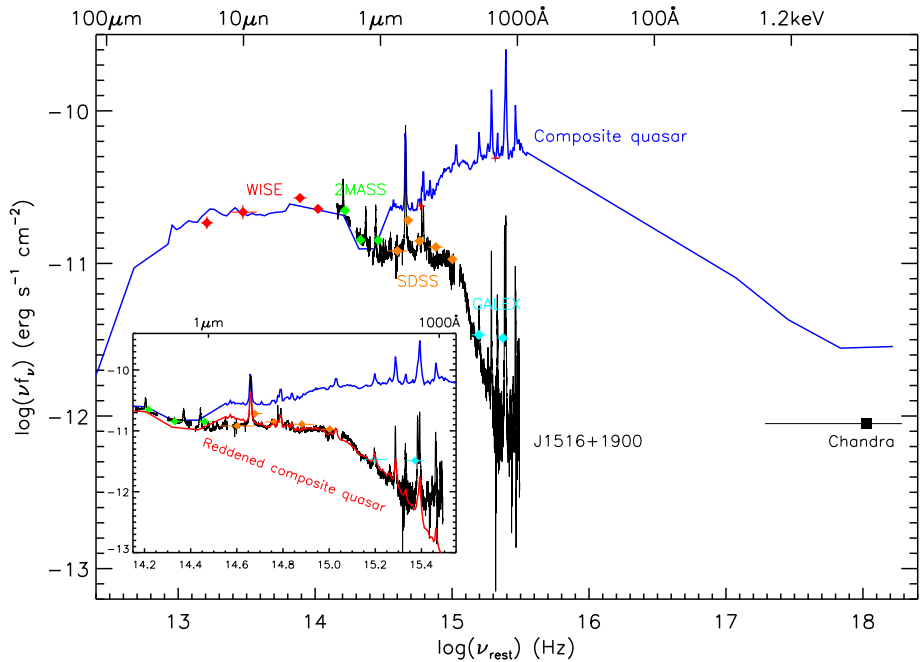}
\caption{Broadband SED of J1516+1900 in the rest frame spanning from infrared to X-ray. The composite quasar spectrum (blue line) normalized at WISE--$W3$ is overplotted for comparison. The observed SED of J1516+1900 is identical to the composite quasar spectrum in the long-ward portion ($\lambda > 1\mu$m), while decreases in the short-ward portion. We zoom in the short wavelength region in the insert panel to make a better demonstration. We redden the quasar composite spectrum (red line) using the SMC extinction curve with an $E(B-V)$ of 0.32 and found it can well match the observed SED of J1516+1900.}
\end{figure*}

With the multi-wavelength photometric and spectroscopic data of J1516+1900, we generate its broadband SED in the rest frame spanning from infrared to X-ray (Figure 1). The spectroscopic data are well consistent with the photometric data, indicating that flux variation is not significant among different observation epochs. As monitored in the Catalina Sky Survey\footnote{http://nesssi.cacr.caltech.edu/DataRelease/} from 2005 July 6 to 2013 June 16, the V-band variations of J1516+1900 are within 0.07 mag, roughly consistent with the median measurement uncertainty of 0.06 mag. Therefor variability of J1705+3543 in optical band is insignificant.

For comparison, we overplot the composite quasar spectrum (Shang et al. 2011) and normalize it to the SED of J1516+1900 at the WISE $W3$ band. In the long wavelength range ($\lambda > 1~\mu m$), the SED of J1516+1900 is consistent with the composite quasar spectrum. In contrast, toward the shorter wavelengths, the continuum flux level of J1516+1900 decreases gradually in the optical band and sharply in the UV band. We zoom in this wavelength range in the insert panel of Figure 1 to better demonstrate this.

We apply a reddening process to the quasar composite spectrum to see if the SED of J1516+1900 can be reproduced by a typical quasar spectrum plus reddening. We use the extinction curves of the Small Magellanic Cloud (SMC). It can be seen that the observed SED of J1516+1900 can be well reproduced, when the composite quasar spectrum is reddened by the SMC extinction curve with an $E(B-V)$ of 0.32. The only mismatch is at the FUV band, where the spectrum of J1516+1900 turns up toward short wavelength range ($\lambda <$ 1200 \AA), implying J1516+1900 may have some extra radiation (e.g., scattered light) contributed to the FUV continuum. Despite this, the good match between the UV-optical SED of J1516+1900 and a reddened quasar suggests that the central continuum source of J1516+1900 is obscured at $E(B-V) \sim 0.32$.

The X-ray emission of J1516+1900 may be obscured even more. When normalized as before, the \textit{Chandra} X-ray emission is much fainter than that of the composite quasar spectrum. To result in such faint \textit{Chandra} X-ray emission, it requires a large hydrogen column density at $N_{\rm H}=5.4 \times 10^{23} ~ \rm cm^{-2}$. However, according to the dust extinction of $E(B-V)=0.32$ derived above, and assuming a SMC gas-to-dust ratio of $N_{\rm H}/E(B-V)=3.7-5.2 \times 10^{22} ~ \rm cm^{-2} ~ mag^{-1}$ (Bouchet et al. 1985), the column density of the obscuring material is estimated to be only $1.18-1.66 \times 10^{22} ~ \rm cm^{-2}$. As SMC has a relative larger gas-to-dust ratio, compared with Milky Way ($N_{\rm H}/E(B-V)=5.0 \times 10^{21} ~ \rm cm^{-2} ~ mag^{-1}$, Burstein \& Heiles, 1978), and Large Magellanic Cloud (LMC, $N_{\rm H}/E(B-V)= 2 \times 10^{22} ~ \rm cm^{-2} ~ mag^{-1}$, Koornneef et al. 1982), the column density should be sufficiently estimated. Therefore the faint \textit{Chandra} X-ray emission of J1516+1900 indicates more absorbing hydrogen atoms located in the front of the obscuring material.

\subsection{Emission Line Spectra}

\begin{figure*}[!htb]
\epsscale{0.8}
\plotone{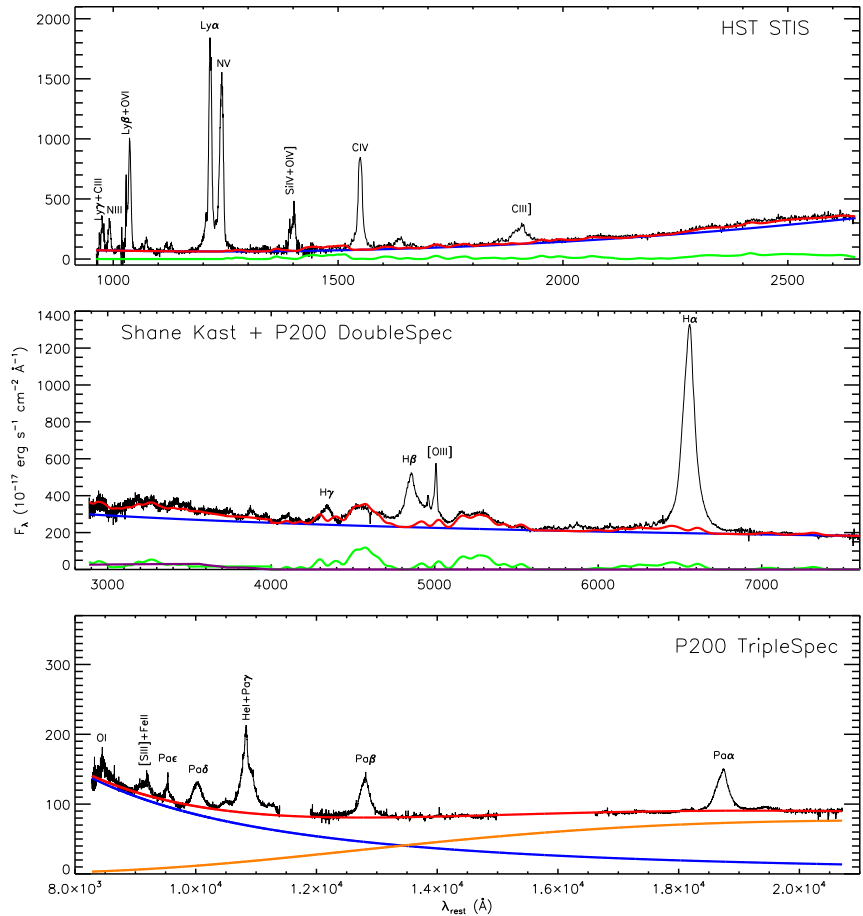}
\caption{Observed UV-Optical-NIR spectra (black) of J1516+1900 in rest-frame wavelengths overlaid with the continuum models (red). \textbf{Top}: UV spectrum taken by \textit{HST}/STIS, and the continuum fitted with a power-law (blue) and a UV iron pseudocontinuum (green). \textbf{Middle}: Optical spectrum combined by Shane DoubleSpec and Hale DoubleSpec spectrum. The optical continuum includes a power law (blue), a Balmer continuum (purple) and an optical iron pseudocontinuum (green). \textbf{Bottom}: NIR spectrum obtained by Hale TripleSpec, and the continuum modelled with a power law (blue) and a black body (orange). Prominent emission lines are labelled in each panel. The UV spectrum shows rich and clear IELs, while the optical and NIR bands are dominated by strong BELs.}
\end{figure*}

Figure 2 presents the overall observed spectra of J1516+1900 from UV through optical to IR. There are rich emission lines shown in all of the bands. BELs lines are very strong shown in the optical and NIR, while become much weaker toward UV and almost disappear in FUV. At the same time, there are rich emission lines with intermediate widths clearly presented in the UV spectra.

To investigate the properties of these emission lines in detail, we subtract the underling continuum form the observed spectra by locally fitting the continuum of the UV, optical, and NIR separately, in waveband windows free from strong emission lines. The continuum modelings of different wavebands are treated slightly differently: (1) We fit the UV continuum using a single power law and UV iron multiplets (Vestergaard \& Wilkes 2001). (2) The optical continuum is fitted with the combination of a power law, a Balmer continuum (supplemented with blended high-order Balmer emission lines), and optical iron multiplets (V{\'e}ron-Cetty et al. 2004). (3) The NIR continuum is fitted using the combination of a power law and a black body. These continuum models are overplotted in Figure 2.

We subtract the continuum models from the observed spectra and display the main emission lines in Figure 3. The UV emission lines (including \lyr, \ciii\ $\lambda$977, \niii\ $\lambda$991, \lyb, \ovi\ $\lambda$1035, \lya, \nv\ $\lambda$1240, \siiv\ $\lambda$1397, \oiv]\ $\lambda$1042, \civ\ $\lambda$1549) are dominated by an intermediate-width component, while their broad component are almost completely absent. More interestingly, in the blue sides of the most strong UV lines of \lya, \nv\ $\lambda$1240, and \civ\ $\lambda$1549, there is also a minor intermediate-width component, blue shifted by $\sim$ 2500 \kmps. We call the major unshifted intermediate-width component as unshifted system, and the minor blue-shifted intermediate-width component as blue-shifted system. The optical and NIR emission lines (such as \hb, \ha, \pb, and \pa) are dominated by their broad components.

We decompose the main emission lines of J1516+1900 into a broad, a narrow, and two intermediate-width components (the unshifted system and the blue-shifted system), to measure the strengths of different emission line components. The method of line decomposition is similar to that described in detail in Li et al. (2015), with small modifications. A single Gaussian is performed to model the narrow and intermediate component, and two Gaussians for the broad component. The same components in different lines are assumed to have the same redshift and line width. For the doublets of \ovi\ $\lambda\lambda$1032, 1037, \nv\ $\lambda\lambda$1239, 1243, \siiv\ $\lambda\lambda$1394, 1403, and \civ\ $\lambda\lambda$1548, 1551, each doublet component is modelled separately with their relative intensity ratios fixed at $1:1$, assuming that the emission is optically thick. The forbidden lines [\oiii] $\lambda\lambda$4959, 5007 are modelled with two components, one narrow component for the line core and one free Gaussian component for the blue wing. An additional broad Gaussian is performed to eliminate the influence of \hb\ ``red shelf'' (Meyers \& Peterson 1985; V{\'e}ron et al. 2002), a red wing extending underneath the [\oiii] double lines. Absorption lines and bad pixels are carefully masked during the fitting process. The best-fit results are shown in Figure 3, and the emission-line parameters are summarized in Table 3.

\begin{figure}[!htb]
\epsscale{1.0}
\plotone{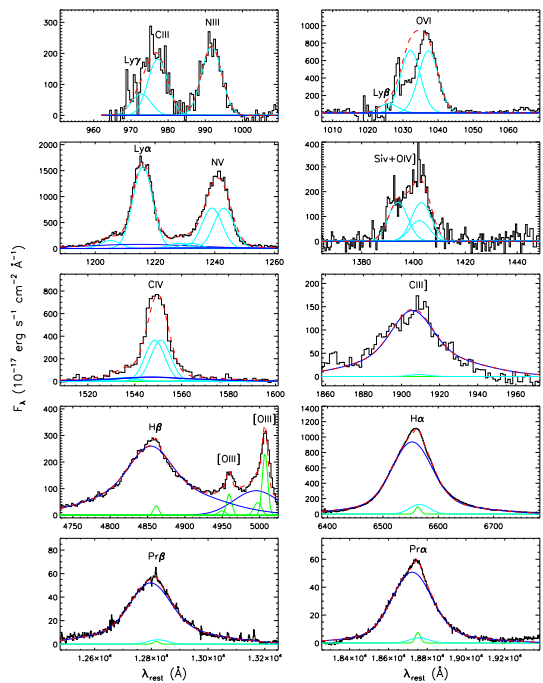}
\caption{Emission lines (black) of J1516+1900 in rest-frame wavelengths sorted from UV through optical to NIR. The UV emission lines are dominated by an intermediate-width component, while their broad component are almost completely absent. In the blue sides of the most strong UV lines of \lya, \nv\ $\lambda$1240, and \civ\ $\lambda$1549, there are also minor intermediate-width components. The optical and NIR emission lines are dominated by their broad components. We decompose these emission lines into the broad (blue), narrow (green), and intermediate-width (cyan) components. The red dashed lines represent the sum of all components.}
\end{figure}

\begin{deluxetable}{lccc}[!htb]
\tabletypesize{\footnotesize}
\tablewidth{0pt}
\tablecaption{Measurements of Emission Lines}
\tablehead{\colhead {} &{BEL} &{IEL(major)} &{IEL(minor)}}
\startdata
Shift (\kmps)&        -551$\pm           5$&          62$\pm           5$&       -2609$\pm          33$\\
\hline
FWHM (\kmps)&        6327$\pm         116$&        1919$\pm          11$&        1919$\pm          11$\\
\hline
Flux ($\rm 10^{-17}~erg~s^{-1}~cm^{-2}$) &&&\\
\lyr&$<          15$&         546$\pm         125$&--\\
\ciii~$\lambda$977&$<          21$&        1443$\pm         118$&--\\
\niii~$\lambda$991&$<          24$&        1431$\pm          83$&--\\
\lyb&$<          63$&         868$\pm         391$&--\\
\ovi~$\lambda$1035&$<         408$&       11986$\pm           0$&--\\
\lya&        4328$\pm         221$&       15517$\pm         196$&        1521$\pm          64$\\
\nv~$\lambda$1240&        1184$\pm           0$&       15650$\pm         106$&        2008$\pm          40$\\
\siiv~$\lambda$1397&$<         849$&        3545$\pm         149$&--\\
\oiv~$\lambda$1402&$<         258$&        1138$\pm         254$&--\\
\civ~$\lambda$1549&        4150$\pm         594$&        9196$\pm         198$&         836$\pm         309$\\
\ciii]~$\lambda$1909&        7140$\pm         272$&          65$\pm          14$&--\\
\hb&       41846$\pm        1496$&           0$\pm           0$&--\\
\ha&      190998$\pm        1474$&        6635$\pm        1306$&--\\
\pb&       15882$\pm         161$&         388$\pm         136$&--\\
\pr&       18903$\pm         215$&         563$\pm         183$&--\\
\enddata
\end{deluxetable}

\subsection{Extinction Analysis For Emission-line Regions}
With the measurements of BELs, we first investigate the extinction for the BELR, using the intensity ratios of BELs in J1516+1900 to those in the composite quasar spectra (Vanden Berk et al. 2001; Scott et al. 2004; Harris et al. 2016). Figure 4 presents the derived intensity ratios of J1516+1900 BELs to composite quasar BEls. The intensity ratios is normalized to unity at \pa. For those heavily blended lines (including \lyr\ and \ciii\ $\lambda$977, \lyb\ and \ovi\ $\lambda$1035), we use the summed intensities (\lyr\ + \ciii\ $\lambda$977, \lyb\ + \ovi\ $\lambda$1035) for both J1516+1900 and the composite spectrum, to reduce the uncertainties of line decompositions. In shorter wavelength range, BELs (such as \lyr, \ciii\ $\lambda$977, \niii\ $\lambda$991, \lyb, and \ovi\ $\lambda$1035) are weak, but they can put strong constraints in estimating the BELR extinction. We use their 3-$\sigma$ upper limits to constrain the extinction. As shown in Figure 4, the intensity ratios gradually decrease toward shorter wavelengths, which can be well modelled by the SMC extinction curve with an $E(B-V)$ of 0.34$\pm0.05$. The dust extinction of BELs is very close to that derived from the broadband SED, implying the BELR and accretion disk is obscured by common dust grains.

On the other hand, the prominent UV IELs shown in the observed spectrum of J1516+1900 indicate that dust extinction is not significant for the IELs. We investigate the IELs extinction through the Lyman line ratios of \lya/\lyr. We use \lyr\ rather than \lyb\ since \lyr\ is more sensitive to the dust extinction. Its measurement is also more reliable. Under the CASE B condition, the intrinsic \lya/\lyr\ is $\sim 1500-2500$ in the low-density limit, and $\sim 30-100$ in the high-density limit (Osterbrock 2006). The observed value in J1516+1900 IELs is $28.4\pm6.5$, roughly consistent with the theoretical ranges of the high-density\hao{-}limit case. Therefore, the IELs should be produced in a region without much dust extinction, as well as having a large gas density.

\begin{figure}[!htb]
\epsscale{1.0}
\plotone{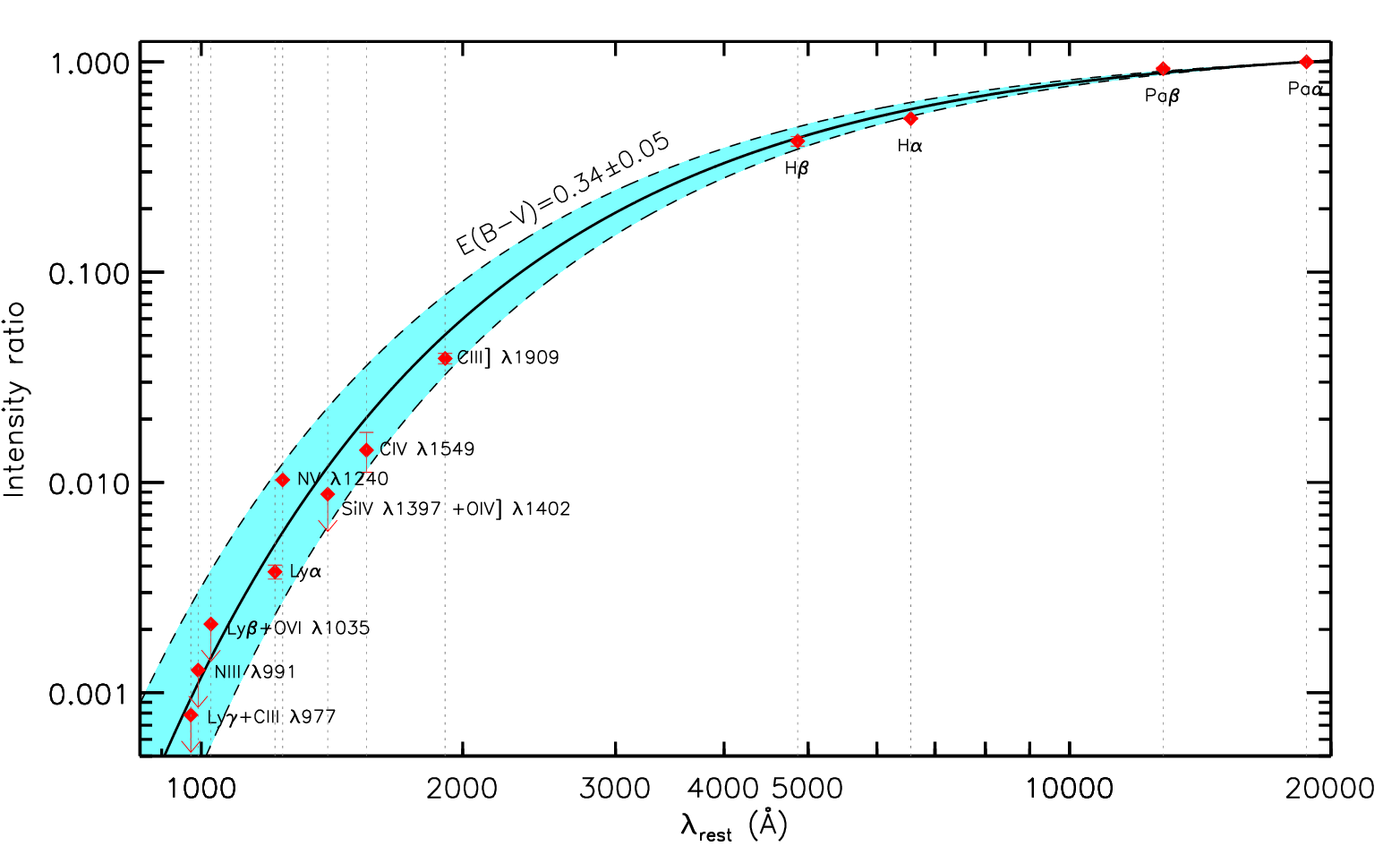}
\caption{BEL intensity ratios (red diamond) of J1516+1900 to composite quasar. The ratios are normalized to unity at \pa. From long to short wavelength lines, the ratios gradually decrease, which is well modelled by the SMC extinction curves with an $E(B-V)$ of 0.34$\pm0.05$.}
\end{figure}

\section{Discussion}
\subsection{AGN Ionization}
Using the measurements for the rich set of IELs in J1516+1900, and combining with the photo-ionization models, we investigate the physical conditions of the IELR. We perform a photo-ionization simulation using the CLOUDY code (Version 13.03, Ferland et al. 1998). We consider a slab-shaped gas that is illuminated by an ionizing source with an AGN SED defined by Mathews \& Ferlan (1987, hereafter MF87). For simplicity, we assume the gas with unique density, homogeneous chemical composition of solar values. The total column density ($N_{\rm H}$) of the gas increases with a grid of $N_{\rm H}$ = $10^{20}~\rm cm^{-2}$, $10^{21}~\rm cm^{-2}$, $10^{22}~\rm cm^{-2}$. For each $N_{\rm H}$, we calculate a two-dimensional grid with variable hydrogen density ($n_{\rm H}$) and ionization parameter ($U$). Since the IELR of this quasar may have a large gas density as mentioned above, we consider a wide $n_{\rm H}$ range from $n_{\rm H}$ = $10^9~\rm cm^{-3}$ to a high value of $n_{\rm H} = 10^{14}~\rm cm^{-3}$. As there are high-ionization IELs in the observed spectra, such as \ovi, \nv, \civ, the IELR may also have a relative high ionization parameter. We restrict the range of $U$ to be from $U=10^{-3}$ to $U=1$. Both of $n_{\rm H}$ and $U$ vary with a step of 0.5 dex.

In total there are three various parameters, $n_{\rm H}$, $U$, and $N_{\rm H}$ in the simulation. We constrain these parameters using the observed IEL ratios measured in J1516+1900. Here we focus on the major IEL system because it has a rich set of IELs that can be used. As shown in Figure 5, for each $N_{\rm H}$, we plot the contours of line intensity ratios (including \lya/\lyr, \civ~$\lambda1549$/\ciii~$\lambda977$, \nv~$\lambda1240$/\niii~$\lambda911$ and \ovi~$\lambda1035$/\oiv]~$\lambda1402$) as functions of $n_{\rm H}$ and $U$. We choose these intensity ratios of two lines from a single element to alleviate ambiguities due to chemical compositions. The colored areas represent the ranges for 1-$\sigma$ measurement errors in this object. When $N_{\rm H}$ increases to $10^{21}~\rm cm^{-2}$, the four intensity ratios suggest the best constrained parameters of $n_{\rm H} = 10^{12.9}~\rm cm^{-3}$, $U = 10^{-1.8}$.

With the constrained $U$ and $n_{\rm H}$, we estimate the distance of the IELR to the ionizing source ($R_{\rm IELR}$). According to the definition of $U$, $R_{\rm IELR}$ can be expressed as $R_{\rm IELR}=\sqrt{ \frac {Q({\rm H})} {4\pi c U n_{\rm H}} } = \sqrt{ \frac {\int_{\nu_0}^{\infty} L_{\nu}/h \nu d \nu} {4\pi c U n_{\rm H}} }$, where $Q(\rm H)$ is the number of photons emitted by the ionization source which can ionize the hydrogen, $L_{\nu}$ is the specific luminosity of the ionizing source, and the integral is over all hydrogen-ionizing photons. Using the dust extinction corrected continuum and assuming an MF87 SED, we have $Q \approx 3.5 \times 10^{56}~\rm photons~s^{-1}$. Once $U$ and $n_{\rm H}$ are given, $R_{\rm IELR}$ can be inferred from the above equation. In Figure 5, we also plot the contours (dotted lines) of $R_{\rm IELR}$ as functions of $n_{\rm H}$ and $U$. With the above constrained $n_{\rm H}$ and $U$, $R_{\rm IELR}$ is found to has a small value of $\sim$ 0.027 pc.

\begin{figure*}[!htb]
\epsscale{1.0}
\plotone{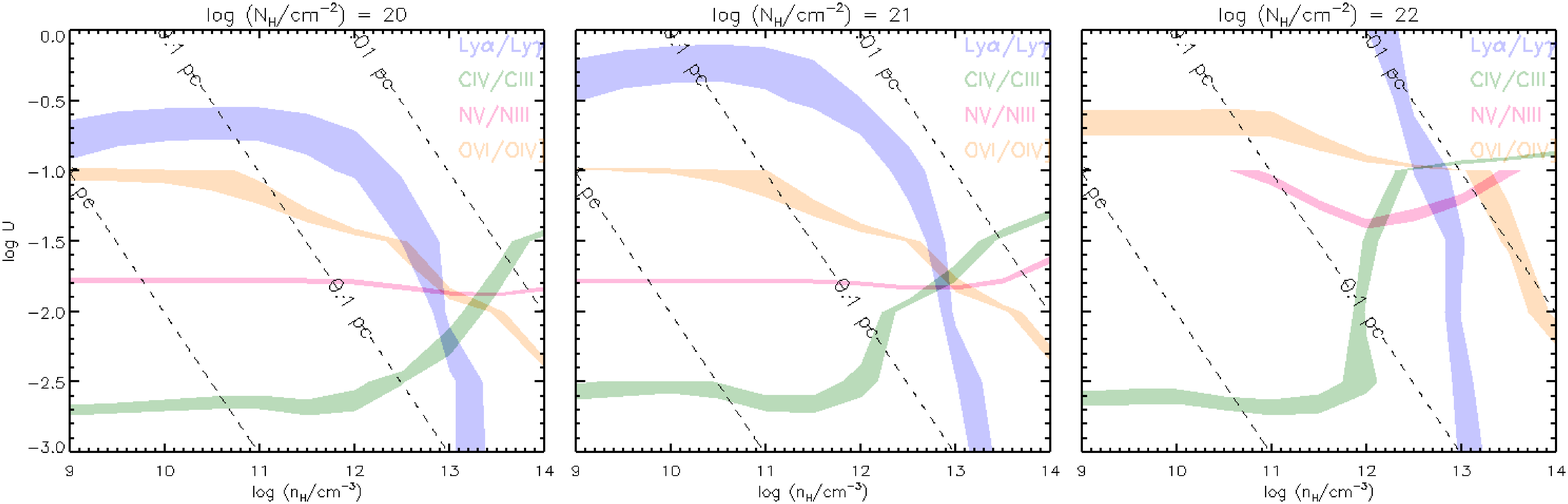}
\caption{Photo-ionization simulations of gas illuminated by an ionizing source with an MF87 SED. We plot the contours of line intensity ratios (\lya/\lyr, \civ/\ciii, \nv/\niii\ and \ovi/\oiv]) as functions of $n_{\rm H}$ and $U$, with increasing $N_{H}$ of $10^{20}~\rm cm^{-2}$ (left), $10^{21}~\rm cm^{-2}$ (middle), and $10^{22}~\rm cm^{-2}$ (right). The filled areas represent the observed ranges of IEL intensity ratios. The four intensity ratios suggest the best parameters of $n_{\rm H} = 10^{12.9}~\rm cm^{-3}$, $U = 10^{-1.8}$, when $N_{\rm H}=10^{21}~\rm cm^{-2}$. The gray dashed lines represent the distance of emission-line region to ionizing source, calculated as $R=(Q({\rm H})/4\pi c U n_{\rm H})^{0.5}$. With the constrained $n_{\rm H}$ and $U$, the distance of the IELR to the ionizing source is derived to be only $\sim 0.027$ pc.}
\end{figure*}

Such a small distance, however, is significantly smaller than the BELR radius of J1516+1900. According to the empirical relation of BELR radius and continuum luminosity given by Bentz et al. (2009), the BELR radius of this quasar is estimated to be $R_{\rm  BELR} \sim 0.17$ pc. Since the BELR of this quasar is suggested to be obscured by dust grains, it is interesting and confusing that the IELR with such a smaller radius is not obscured.

A possible mechanism of avoiding materials in the inner accretion disk to be obscured might be the ``failed wind'' model (e.g., Proga \& Kallman 2004; Sim et al. 2010). In the inner region of the accretion disk, the material is launched from the disk plane driven by local disk UV line radiation, but not accelerated to reach escape speed due to overionization by central X-ray radiation. The wind eventually falls back to the disk plane. This scenario results in a ``failed wind'', which has a quite high density at large scale heights above the disk plane (Proga et al. 2005). Besides, the magnetic field in the inner disk can help to stir the material up to even greater scale heights, and result in a larger covering factor (Schurch 2006). This model allows the ``failed wind'' to be observed, even though the disk UV$-$optical radiation region and the BELR are obscured in the observer's light of sight. The ``failed wind'' blocks the central X-ray and extreme UV radiation and prevents the outer wind from being overionized, as well as itself ionized and can produces emission lines through photo-ionization processes. However, the velocity dispersion of the ``failed wind'' is expected to be very large, since the ``failed wind'' locates in the inner accretion disk. The rotation velocity in the radius of IELR is estimated to be $\sim 10^4$ \kmps, much larger than the observed IEL velocity. Therefore, it is unlikely to result in emission lines with intermediate-widths through the ``failed wind'' model.
\subsection{Shock Ionization}
A reasonable possibility maybe that the ionizing source of the IELs does not locate in the central accretion disk, but in an outer region beyond the obscuring materials. In this case, $R_{\rm IELR}$ calculated by the photo-ionization model should not be interpreted as the distance of IELR to the central region. Meanwhile, the small radius of IELR to the ionizing source suggests that the IELR may also locate in an outer region beyond the obscuring materials, which is to say that the IELR can not be obscured in the observer's line of sight.

The gas density in AGNs is generally believed to decrease from central to outer region. Usually, the accretion disk and the BELR has a high gas densities in the range of $\sim 10^{9-13}~\rm cm^{-3}$. The inner surface of the dusty torus has a gas density of $\sim 10^{6-9}~\rm cm^{-3}$. In the NELR and the interstellar medium region, the gas densities are only $\sim 10^{2-3}~\rm cm^{-3}$ (e.g., Peterson 1997). To produce $\sim 10^{13}~\rm cm^{-3}$ ultradense gases in an outer region beyond the obscuring materials, some extra physical mechanism should be required. Shock is a common mechanism to cause high-density gases in astrophysics. At the scale of the dusty torus, the outflow in quasars can collide onto some materials, which may result in high-density and high-temperature ionized gases.

\begin{figure*}[!htb]
\epsscale{1.0}
\plotone{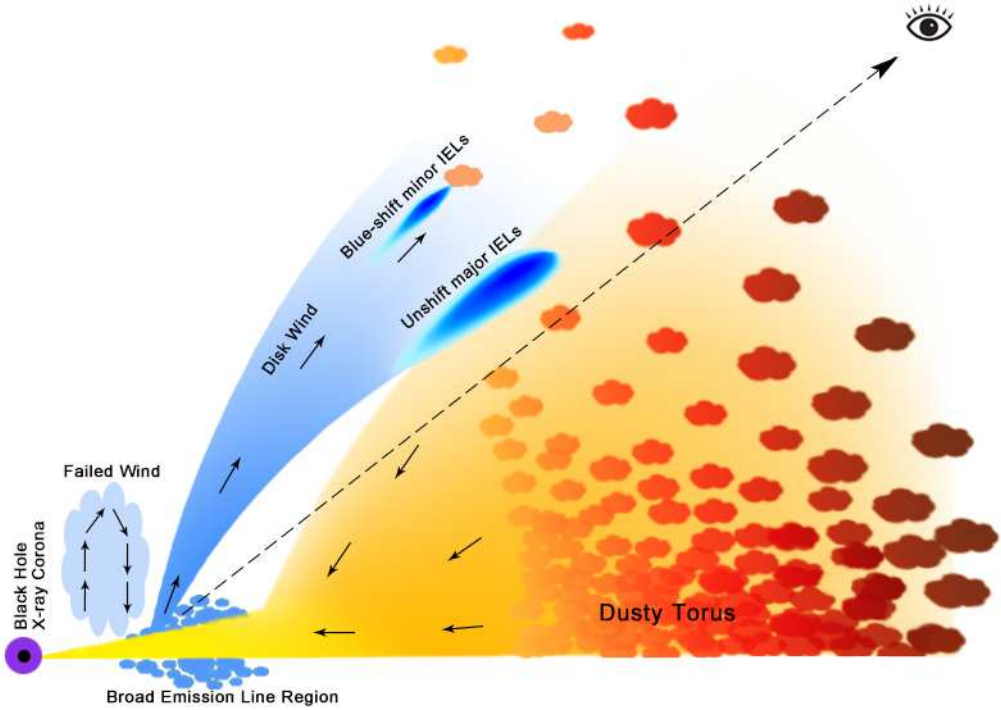}
\caption{A schematic diagram for the IEL origin. The light of sight to the central accretion disk and BELR are obscured by dusty material near the boundary of the dusty torus, which results in the reddened SED and reddened BELs. The outflow gases, launched from the accretion disk with high kinetic energy, collides onto the inner wall of the dusty torus and shock-heat the gases to cause the unshift major IELs. The outflowing gases may also collide into surrounding isolated clouds, and give rise to blue-shifted minor IELs.}
\end{figure*}

\begin{figure*}[!htb]
\epsscale{1.0}
\plotone{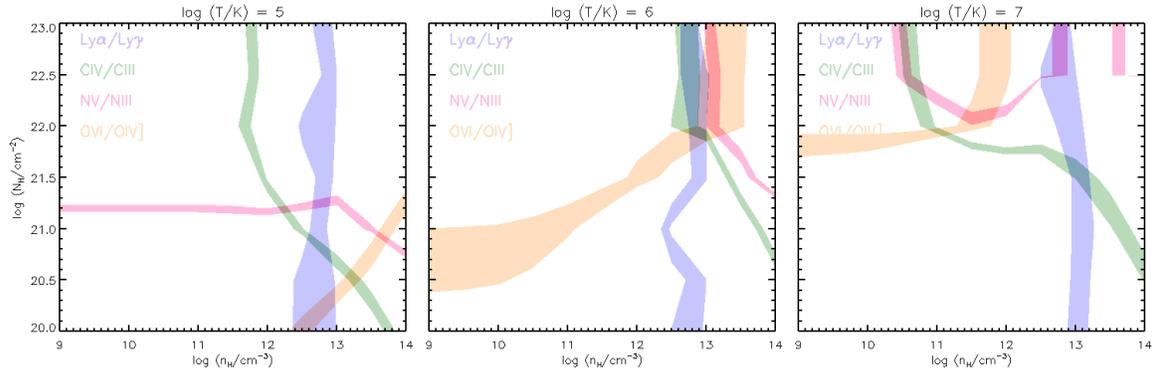}
\caption{Photo-ionization simulations of gas heated by a shock with increasing $T$ of $10^5$ K (left), $10^6$ K (middle), and $10^7$ K (right). For each temperature, we plot the contours of line intensity ratios as functions of $n_{\rm H}$ and $N_{\rm H}$. The filled areas represent the observed ranges of IEL intensity ratios. The four intensity ratios give the best parameters of $n_{\rm H} \sim 10^{13}~\rm cm^{-3}$, $N_{\rm H} \geq 10^{22}~\rm cm^{-2}$, when $T=10^6$ K.}
\end{figure*}

Figure 6 displays a cartoon to illustrate this shock scenario. The light of sight to the central accretion disk and BELR are obscured by dusty grains near the boundary of the dusty torus, which results in the reddened SED and reddened BELs observed in J1516+1900. The outflow is launch from the inner accretion disk and accelerated by disk UV line radiation. When arrives at the region of dusty torus, the outflow accumulates high kinetic energy and collides with materials, which results in high-density and high-temperature ionized gases and produce emission lines with intermediate widths. If the outflow collides onto the inner wall of the dusty torus, the outflow momentum might be almost lost through a violent collision and cause the unshifted major IELs. If the outflow collides onto surrounding isolated clouds, a fraction of its momentum could be retained through a slight collision, which gives rise to the blue-shifted minor IELs.

To check this supposition, we also perform a CLOUDY simulation for the shock model. The simulation settings for our shock model are similar to Zhang et al. (2019), which described a CLOUDY simulation for high-density shock in detail. As the radiation in the cooling zone of the photo-ionizing shock is dominated by the thermal bremsstrahlung radiation, the incident radiation in our shock simulation adopts the model of Sutherland et al. (1993), which evaluated an ionizing spectrum of thermal bremsstrahlung radiation. The ionizing spectrum is characterized by the temperate ($T$). We consider a possible temperate range of $10^5<T<10^7~\rm K$, in which the ionizing spectrum, unlike the AGN SED, peaks at extreme UV to soft X-ray band, and falls rapidly at higher and lower energy. The other simulation settings are same as the AGN model described above. Figure 7 shows the simulation results. The shock model can reproduce the four considered IELs ratios measured in J1516+1900 under the parameters of $n_{\rm H} \sim 10^{13}~\rm cm^{-3}$, $N_{\rm H} \geq 10^{22}~\rm cm^{-2}$, in the case of $T=10^6$ K.

Adopting this parameters derived above, we use CLOUDY code and export the intensities for all of the UV IELs (including \lyr, \ciii\ $\lambda$977, \niii\ $\lambda$991, \lyb, \ovi\ $\lambda$1035, \lya, \nv\ $\lambda$1240, \siiv\ $\lambda$1397, \oiv]\ $\lambda$1042, \civ\ $\lambda$1549, \ciii]\ $\lambda$1909) in J1516+1900. Figure 8 shows the comparison between the shock model predicted intensities and the observed IEL intensities (both are normalized to \lya\ intensity). The UV IEL intensities observed in J1516+1900 are consistent with those predicted by the shock model.

\begin{figure}[!htb]
\epsscale{1.0}
\plotone{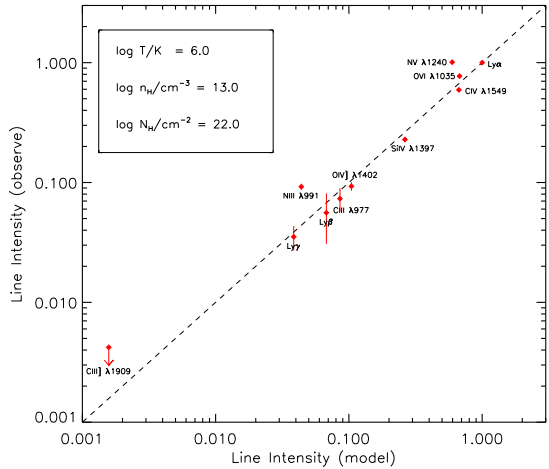}
\caption{Comparison between the shock model predicted line intensities and the observed line intensities for all of the UV IELs in J1516+1900. Both of the model predicted and observed line intensities are normalized by \lya. The dashed line represents a 1:1 relationship.}
\end{figure}

\section{Summary and Implication}
We presented a detailed analysis of the partially obscured quasar J1516+1900 with its archived data and follow-up observations. Both of its broadband SED and BELs are reddened by an SMC-like extinction curve, while its UV spectrum shows rich and clear IELs with FWHM about 1900 \kmps\ in two systems, an unshifted major one, and a blue-shifted minor one.

With the aids of the observed IELs and assuming the IELs are photo-ionized by the central accretion disk, the distance of the IELR to the ionizing source is estimated to be only $\sim$ 0.027 pc. Such a distance is significantly smaller than the BELR radius of $\sim$ 0.17 pc. This is contradict with the IEL observational features of intermediate widths and no significant dust extinction. The ``failed wind'' model might avoid the gases to be obscured, but unlikely to result in emission lines with intermediate widths.

A reasonable explanation for the IEL origin is that the IELs are produced by shock-heated high-density and high-temperature gases, rather than photo-ionized by the central accretion disk. The outflow, launched from the accretion disk with high kinetic energy, collides onto the inner wall of the dusty torus and shock-heat the gases to cause the major emission lines. The outflowing gases may also collide onto surrounding isolated clouds, and give rise to blue-shifted minor emission lines.

If the IELs are indeed reproduced by the shock model, the IELs observed in quasars might provide a new avenue to understand the physical processes of quasar outflow. For example, the IEL intensities should relate to the outflow kinetic energy. The intensity ratios of IELs and BELs could increase with the increasing interactions between outflow gases and materials at the scale of the dusty tours. It can be expected that for some quasars with extreme outflow, their emission lines could be dominated by the IELs, even though the BELs are not obscured. Such extreme IELs are well worth studying the quasar outflow and will be discussed in detail in our future works.

We are grateful to Vestergaard for kindly providing us the iron emission line templates. 

Some of the data presented in this paper were obtained through the Telescope Access Program (TAP), which has been funded by the Strategic Priority Research Program ``The Emergence of Cosmological Structures'' (XDB 09000000), National Astronomical Observatories, Chinese Academy of Sciences, and Special Fund for Astronomy from the Ministry of Finance.

This research is based on observations made with the NASA/ESA Hubble Space Telescope obtained from the Space Telescope Science Institute, which is operated by the Association of Universities for Research in Astronomy, Inc., under NASA contract NAS 5–26555. These observations are associated with program 9161. This publication makes use of data products from the Two Micron All Sky Survey, which is a joint project of the University of Massachusetts and the Infrared Processing and Analysis Center/California Institute of Technology, funded by the National Aeronautics and Space Administration and the National Science Foundation. Also, this publication makes use of data products from the WISE, which is a joint project of the University of California, Los Angeles, and the Jet Propulsion Laboratory/California Institute of Technology, funded by NASA.

Funding for SDSS and SDSS-II has been provided by the Alfred P. Sloan Foundation, Participating Institutions, National Science Foundation, U.S. Department of Energy, NASA, Japanese Monbukagakusho, Max Planck Society, and Higher Education Funding Council for England. The SDSS is http://www.sdss.org/.

SDSS is managed by the Astrophysical Research Consortium for the Participating Institutions. The Participating Institutions are the American Museum of Natural History, Astrophysical Institute Potsdam, University of Basel, University of Cambridge, Case Western Reserve University, University of Chicago, Drexel University, Fermilab, Institute for Advanced Study, Japan Participation Group, Johns Hopkins University, Joint Institute for Nuclear Astrophysics, Kavli Institute for Particle Astrophysics and Cosmology, Korean Scientist Group, Chinese Academy of Sciences (LAMOST), Los Alamos National Laboratory, Max-Planck-Institute for Astronomy (MPIA), Max-Planck-Institute for Astrophysics (MPA), New Mexico State University, Ohio State University, University of Pittsburgh, University of Portsmouth, Princeton University, United States Naval Observatory, and the University of Washington.

\end{document}